# Structured electrode additive manufacturing for lithium-ion batteries

Soyeon Park[1], Kaiyue Deng[1], Baohui Shi[1], Yuanyuan Shang[1], Kun (Kelvin) Fu[1,2, *]

[1]Department of Mechanical Engineering, University of Delaware, Newark, DE 19716, USA
[2]Center for Composite Materials, University of Delaware, Newark, DE 19716, USA

[*] Corresponding authors

Emails: kfu@udel.edu

**Abstract:** A thick electrode with high areal capacity has been developed as a strategy for high-energy-density lithium-ion batteries, but thick electrodes have difficulties in manufacturing and limitations in ion transport. Here, we reported a new manufacturing approach for ultra-thick electrode with aligned structure, called structure electrode additive manufacturing or SEAM, which aligns active materials to the through-thicknesses direction of electrodes using shear flow and a designed printing path. The ultra-thick electrodes with high loading of active materials, low tortuous structure, and good structure stability resulting from a simple and scalable SEAM lead to rapid ion transport and fast electrolyte infusion, delivering a higher areal capacity than slurry-casted thick electrodes. SEAM shows strengths in design flexibility and scalability, which allows the production of practical high energy/power density structure electrodes.

## 1. Introduction

Lithium-ion batteries (LIBs) are essential as a functional energy source for electric vehicles (EVs) as the demand to reduce carbon emissions from transportation increases. Currently, the United States aimed to produce half of all new car sales as electric vehicles by 2030[1], and annual LIB development is projected to increase about 8-fold over the next decade, reaching nearly 2 TWh of capacity worldwide[2]. To achieve LIB development targets, there is a strong demand for improved LIB performance at a lower cost (< $80-100/kWh), such as fast charging capabilities (<15 minutes) and high energy density (range of EVs to 300 miles).[3] Current slurry-casted electrodes have limited thickness and simple 2D structure, and difficulties in minimizing the ratio of inactive materials while maintaining LIB performance per mass or volume of active material. In addition, manufacturing processes that include coating and drying steps have inherent challenges in producing high thickness due to the cracking and delamination of electrode films from current collectors during drying. Therefore, there is an opportunity to improve LIB performance and producibility by adopting structured electrode manufacturing processes and scaling up.

There have been trials to increase the thickness of 2D slurry-casted electrodes, but thick 2D structure causes the sluggish ion transport across the thick electrodes, which is an obstacle to achieving high power. Furthermore, new approaches to reducing tortuosity for rapid ion transport by creating through-thickness holes using laser[4,5] and punchers[6] or by making aligned structures using advanced techniques such as magnetic fields[7,8], freeze-dry[9,10], and vapor deposition[11] have been reported. These new fabrication techniques for low-tortuosity electrodes have enhanced the ion movement, whereas making pathways with mechanical tools still holds limitations in further increasing thickness of electrodes due to the use of slurry, and new processes employing additional fabrication system seem quite challenging and difficult to scale beyond lab research. In recent times, 3D printing has been showing a great potential to fabricate electrodes with not only high thickness and low tortuosity but also complex geometry. Attempts have been made to produce structure electrodes using ink-based printing[12,13]. However, the structure made of the ink becomes unstable and tends to collapse under gravity as subsequent layers are printed on top preventing further increase in thickness and improving structural strength and structure geometry.

Here, we developed a novel structure electrode manufacturing approach, called Structured Electrode Additive Manufacturing (SEAM), which can orient active materials in a though-thickness direction, to produce electrodes of high thickness and low tortuosity. SEAM induces the shelf-alignment of the electrode material by the shear flow that occurs while extruding the molten state of the polymer-active material composite, called Shear-aligned Polymer Extrusion (ShAPE) technology. These thick electrodes with aligned structures enable a fast ion transport in the infiltrated liquid electrolyte, resulting in high energy density and high power of LIB. SEAM has several unique features in the fabrication of structured electrodes, such as design flexibility, scalability, high throughput, and safety due to the solvent-free approach. In addition, the printed structured electrodes have high compressive strength while maintaining a low tortuosity. The schematic of SEAM using SHAPE technology is shown in Figure 1. When a 3D printer manufactures structured electrodes, shear generated within the nozzle provides alignment of graphite, used as active material, along with the direction of the extrusion. The printing direction was modified in order to align the printed graphite in the through-thickness direction of the electrode. The ordered structure of graphite not only provides a short pathway for fast ion transport within the thick electrode, but also allows the ions to intercalate into the graphite easily.

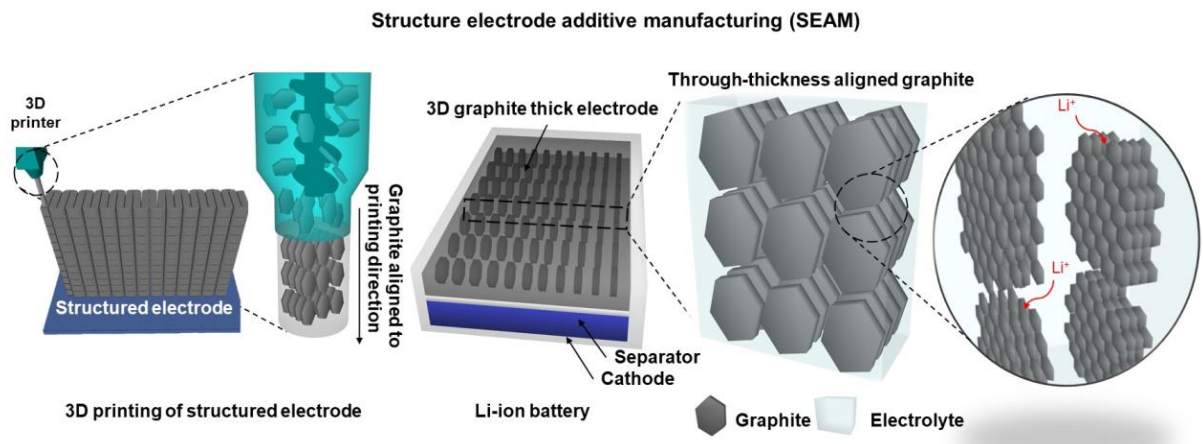

**Figure 1**. Schematic of structure electrode additive manufacturing (SEAM) for 3D graphite thick electrode with low tortuous structure. Shear-aligned polymer extrusion (SHAPE) technology in SEAM orients graphite aligned with a printing direction. Through-thickness aligned graphite allows the fast Li-ion transport and fast electrolyte infusion within the thick anode.

## 2. Results and Discussion

Figure 2 shows a 3D printing process of the aligned graphite 3D electrodes using the customized graphite composite pellets and Pellet Mini 3D printer. As shown in Figure 2a, fabrication for structure electrodes begins with preparing graphite composite pellets and printing the electrodes in specific direction to remove polymer in inert gas. In the preparation of the composite feedstocks, graphite, polylactic acid (PLA), and carbon additives (carbon black (CB) and multi-walled carbon nanotubes (MWCNTs)) were mixed with a weight ratio of 21/70/9. The CB and MWCNTs were used to improve the electrical conductivity and mechanical strength of the 3D graphite electrodes in the printing direction, respectively. The graphite composites were chopped into small pieces to feed the 3D printer, as shown in Figure 2b. Our composite feedstock contains 30 wt.% graphite and carbon additives, and this carbon contents is higher than commercial graphite filaments with a lower load of around 8%. The ratio of carbon components (graphite and carbon additives) in the composite pellets was confirmed through thermal analysis, showing 69.9 wt.% of PLA and 29.1 wt.% of carbon components (Figure 2c). Raman spectroscopy was also implemented to identify the existence of graphite in the composite pellets. To compare with pure graphite powders, PLA in the composite feedstocks was removed by carbonization at 800 °C in the Nitrogen. As shown in Figure 2d, both the carbonized graphite composite pellets and graphite powders showed high intensity at 1355 cm$^{-1}$ and 1586 cm$^{-1}$, which are D band and G band, respectively, and weaker intensity at 2695 cm$^{-1}$. These peaks represent typical defected graphite in the Raman spectrum.[14] Compared to the graphite powders, the composite pellets had a wider band between 1355 cm$^{-1}$ and 1586 cm$^{-1}$ due to the overlapping of the carbon additive peaks, which increase disorder.[14] Using the fabricated composite pellets, the electrodes were printed with a Pellet Mini 3D printer (Figure 2e). Using SHAPE technology applied in SEAM, the alignment of the graphite with respect to the printing direction was achieved, which is driven by the shear forces generated when the molten pellets pass through a small nozzle.[15,16] The thick graphite electrodes with high loading of graphite and aligned structure were successfully printed without defects such as warping, delamination, and layer shifting (Figure 2f). A specific pattern of the electrode was printed to match the printing direction and the thickness direction of the electrode, and as-printed electrodes showed a dense structure by following the printing direction well (Figure 2g).

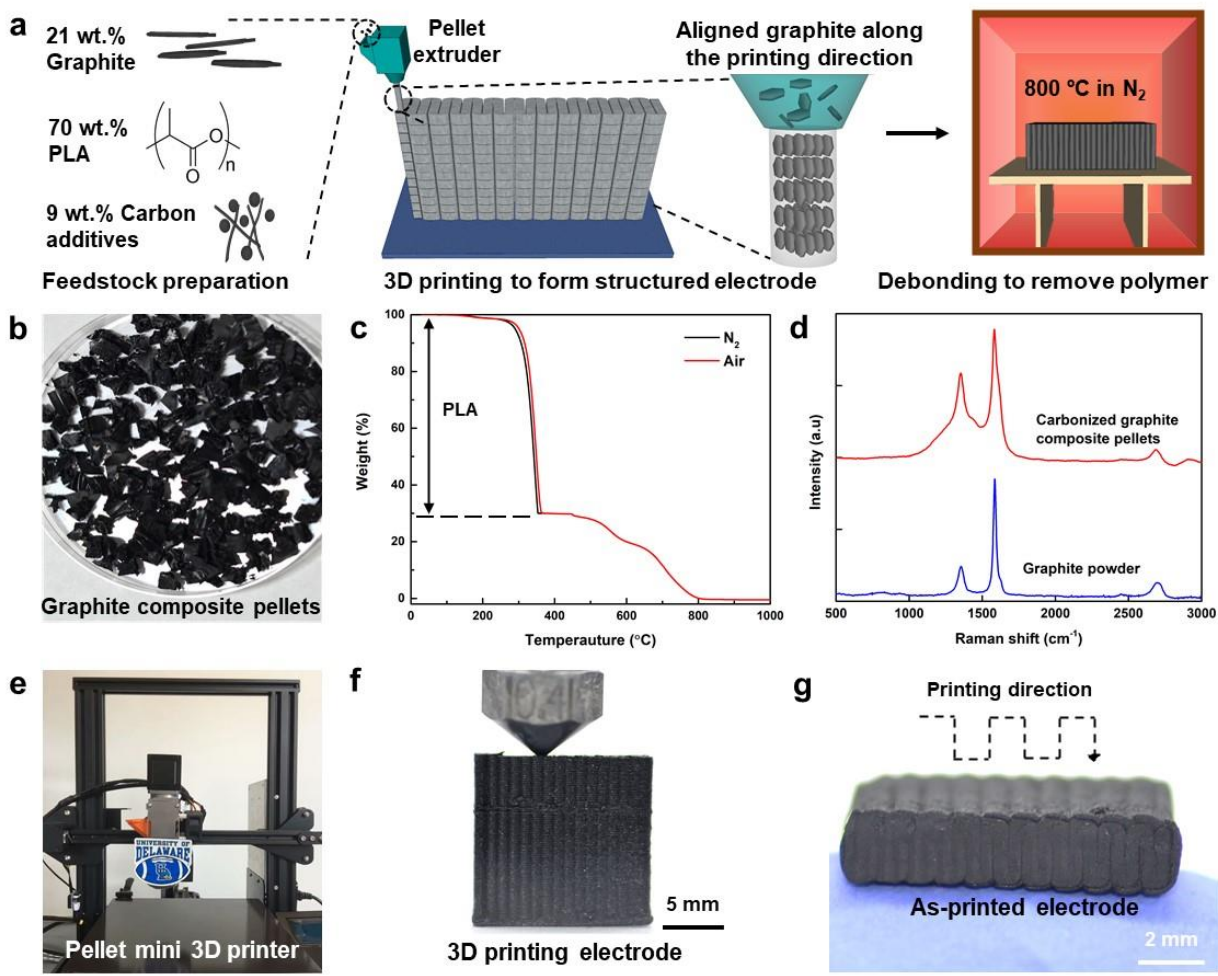

**Figure 2**. The 3D printing process for structured electrodes. (a) 3D graphite electrode fabrication process from 3D printing of 3D structure to polymer (polylactic acid, PLA) removal. Using composite pellets (Graphite, PLA, carbon additives), thick electrodes were printed, and polymer carbonization was carried out. (b) Photo image of the graphite composite feedstock. (c) Thermogravimetric analysis of carbon composite filaments at nitrogen gas until reaching 450 °C, holding isothermal conditions for 30 minutes, and then changing the gas with air. (d) Raman spectra of the carbonized carbon composite filaments and graphite powder. (e) Photo image of pellet mini 3D printer. (f) Photo image of 3D printing of the 3D graphite electrode. (g) Photo image of the printed electrode on the side view showing a dense and aligned structure according to the printing direction.

Prior to using the as-printed samples as electrodes, the upper and lower surface of the samples were polished, and PLA was carbonized at 800 ºC in nitrogen gas to remove polymer and create a through-thickness path for Li-ions to pass through the thick electrodes. The removal of the polymer did not affect the structure of the printed objects, but there were dimensional changes in the area and thickness about 23.4% and 9.5%, respectively (Figure 3a). Even after removing polymers from the samples, the 3D graphite electrodes showed good structural stability, as proved by the mechanical tests (Figure 3b). Our 3D graphite electrodes exhibited a higher compressive modulus of 5MPa and a strength of 7.1MPa than other 3D printed carbon samples (e.g., graphene oxide (GO), reduced graphene oxide (rGO), and graphite-cellulose nanofibers (G-CNF)) due to the ordered graphite structure.

These 3D graphite electrodes were attached to the copper foil to ensure good contact and high electrical conductivity between them, as shown in Figure 3c. Also, parallel stacking of the 3D graphite electrodes on the same current collector could be implemented to produce a high energy density. The fabrication process of the double-sided 3D graphite electrodes on a current collector is schematically shown in Figure 3d. As-printed electrodes were attached to a porous copper foil using heat and then carbonized in an inert gas. The porous current collector allows the Li-ion to pass through a porous current collector.

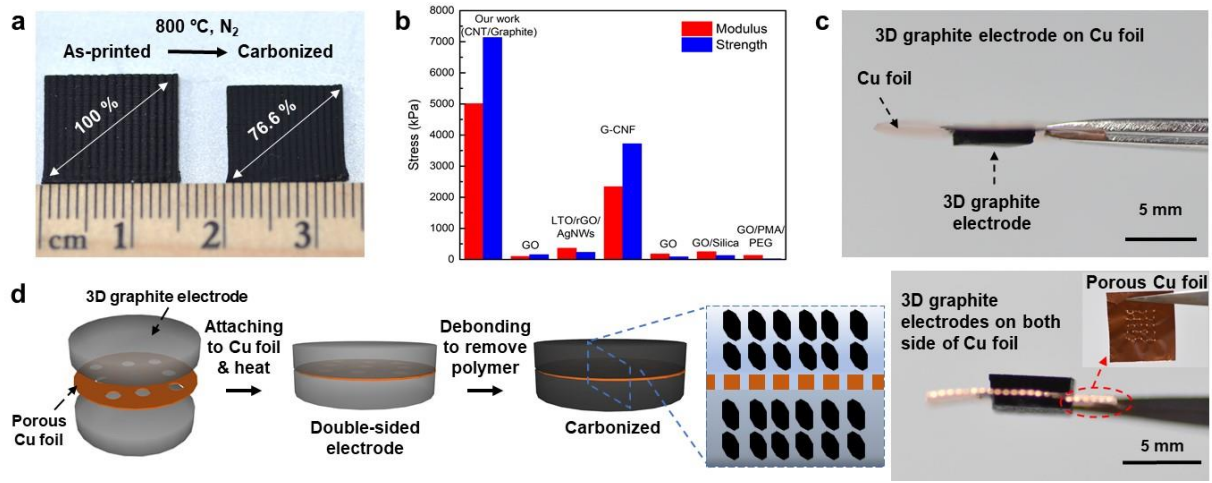

**Figure 3**. Fabrication of structure electrodes with as-printed electrodes. (a) Photo image of comparing dimensional changes of the printed samples before and after PLA removal. (b) Compressive strength and modulus of the 3D graphite electrodes and other 3D printed electrodes. (c) Photo image of the 3D graphite electrode pasted on the copper current collector. (d) Photo image of the 3D graphite electrodes passed on both sides of a porous copper current collector.

The graphite alignment within the electrode was assessed by various characterization tests such as scanning electron microscopy (SEM), X-ray diffraction (XRD), and nano-computed tomography (nano-CT) (Figure 4). The 3D graphite electrode showed a clear printing path in both top and side views in low-magnified SEM images. The magnified SEM images showed that most graphite tended to align along the printing direction, which is out-of-plane at the top (X-Y plane) and vertically aligned when viewed from the side (Y-Z plane) (Figure 4a). The aligned graphite will provide a straight ion transport pathway within the thick electrode, allowing rapid ion transport in the infiltrated liquid electrolyte. In addition, XRD data offered information about the graphite arrangement within the 3D electrodes, showing the same tendency as the SEM images (Figure 4b). XRD tests were conducted in different orientations of the 3D graphite electrodes (horizontal and aligned (printing) directions) to probe the arrangement of the graphite. The alignment degree of graphite was confirmed by tracking the intensity of the (002) plane, corresponding to the graphite basal plane.[8] Compared to the aligned samples, the horizontal samples exhibited about 2 times higher peaks in the (002) plane at 26°, suggesting that the graphite was oriented along the printing direction. The weaker peaks at 44°, 77°, and 83° marked as star symbols in the graph correspond to the other carbon contents, including MWCNTs and carbon black. Furthermore, to quantify the arrangement of the graphite, nano-CT was also implemented. As shown in Figure 4c, the graphite structure of the 3D graphite electrode was compared with a reference sample made by the slurry casting of equal content ratios. In the reference sample, graphite was randomly distributed, whereas the 3D printed sample had aligned graphite along the printing direction. These characterization tests demonstrated that 3D printing using SHAPE technique could fabricate ordered structures of nanomaterials such as graphite along the printing direction.

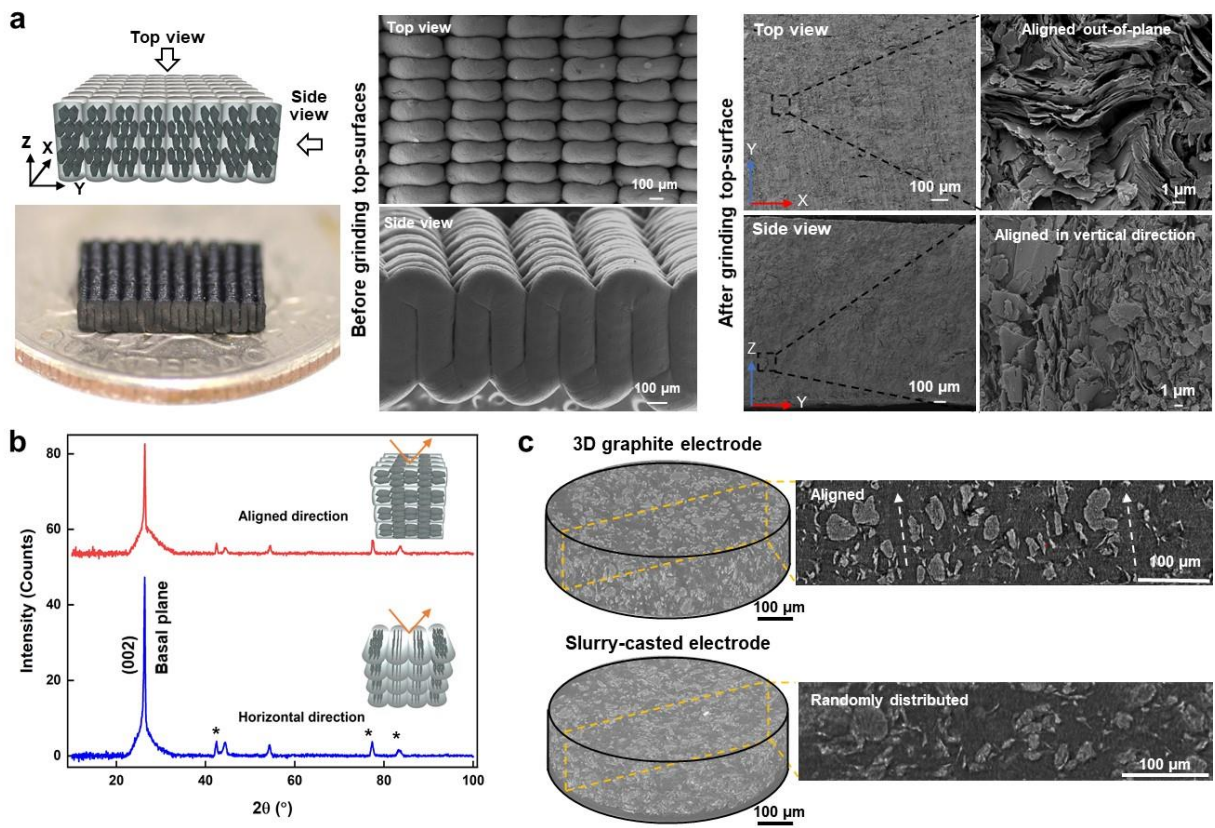

**Figure 4**. Morphological and structural characterization of the 3D graphite electrode. (a) Scanning electron microscopy (SEM) image of 3D graphite electrode. Most of the graphite was aligned out-of-plane in the top view, and graphite tended to be aligned along the printing direction from the side view. (b) X-ray diffractograms (XRD) of the 3D graphite electrode in different directions (horizontal and aligned directions). The graphite showed the aligned structure following the printed direction. The star symbols in the graph represent the existence of carbon additives. (c) Nano-computed tomography (nano-CT) images for the slurry-casted electrode and 3D graphite electrode. Compared to the reference samples, the printed electrodes have ordered graphite structure, leading to fast ion movement in the thick electrodes.

The electrode chemical performance of the 3D graphite thick electrode was tested in a half-cell configuration. To show the enhanced transport of Li-ion due to the aligned structure of the graphite in comparison to the slurry-casted electrode, rate capability tests were performed on both samples (Figure 5). The 3D thick graphite electrodes with different thicknesses (0.5 mm and 1mm) were made, and they had an areal loading of 16.44 mg/cm$^2$ and 46.67 mg/cm$^2$, respectively. The slurry-casting electrode with 1 mm thickness was used as a reference sample and had the same composition and an areal loading of 38.86 mg/cm$^2$. Figures 5a, b, and c give the charge/discharge voltage profiles of the 3D graphite electrode and the slurry-casted electrode at different current densities (0.1- 1C). The 3D graphite electrode with 0.5 mm delivered the highest specific capacity of 442.0, 165.4 and 82.6 mAh/g at 0.1, 0.5, and 1C (where 1C= 6.12 mA/cm$^2$), respectively. In comparison, the 3D graphite electrode with 1 mm showed slightly lower specific capacity of 393.8, 148.7, and 77.3 mAh/g at 0.1, 0.5, and 1C (where 1C= 17.36 mA/cm$^2$) compared to the thinner 3D graphite electrode but had around two times higher areal capacity of 18.4, 6.9 and 3.6 mAh/cm$^2$ at 0.1, 0.5, and 1C due to the higher thickness. These electrochemical performances of 3D graphite electrodes were superior to that of the slurry cast thick electrodes, which delivered a very lower specific capacity of 292.8, 2.0, and 25.5 mAh/g at 0.1, 0.5, and 1C (where 1C= 14.45 mA/cm$^2$), respectively. The overpotential was measured, and the values were calculated based on the potential difference between approximately 0.1V versus Li$^+$/Li potential plateau at 0.1C and the corresponding potential plateau at each elevated rate (Figure 5d). As the current increased, the gap between the overpotentials of the 3D graphite electrode and the slurry cast electrode increased, more than doubling at 1C. The improved electrochemical performance of the 3D graphite sample was also observed in the electrochemical impedance spectroscopy (EIS) tests. The semicircle in the graph means the charge-transfer resistance at the high-frequency region. In Figure 5e, the 3D graphite electrode showed lower charge-transfer resistance of 71.03 ohm, while the slurry-casted sample had 99.40 ohm. These results were attributed to the aligned structure of graphite in the through-thickness direction of the electrode, which leads to the shortest transfer path and the faster

Li-ion movement within the electrode. Lastly, the specific capacities of 3D graphite and reference electrode as a function of areal current density were compared, and both were measured at each 0.1, 0.2, 0.5, and 1C (Figure 5f). At the same thickness of about 1 mm, the 3D graphite electrode exhibited higher capacity than the slurry-casted sample at all rates, showing superior rate performance. All electrochemical results described the importance of active material structure for enhanced performance at high rates with high areal loading. Compared to the reported 3D printed thick electrode using ink or low contents of composite filaments, our 3D graphite thick electrodes produced by SEAM had high loading of active materials as well as exhibited the highest specific capacity at the same areal current density (Figure 5g). These results demonstrate the potential of 3D printing of thick electrodes for high energy density and high-power LIBs that can be achieved with a simple and safe fabrication methods.

3D printing of electrodes has another advantage compared to slurry-casting or other advanced manufacturing methods (e.g., magnetic fields, freeze-dry, and vapor deposition) in terms of design flexibility and scalability. In Figure 5h, the interdigitated structure was fabricated, which is a prominent design for increasing the contact area of two sides of electrodes by reducing ion transfer distances between the electrodes.[17] Also, as shown in Figure 5i, 3D printing has almost no limit to the size of the electrode, and in this experiment, a large electrode with a size of 5×5 cm$^2$ was obtained.

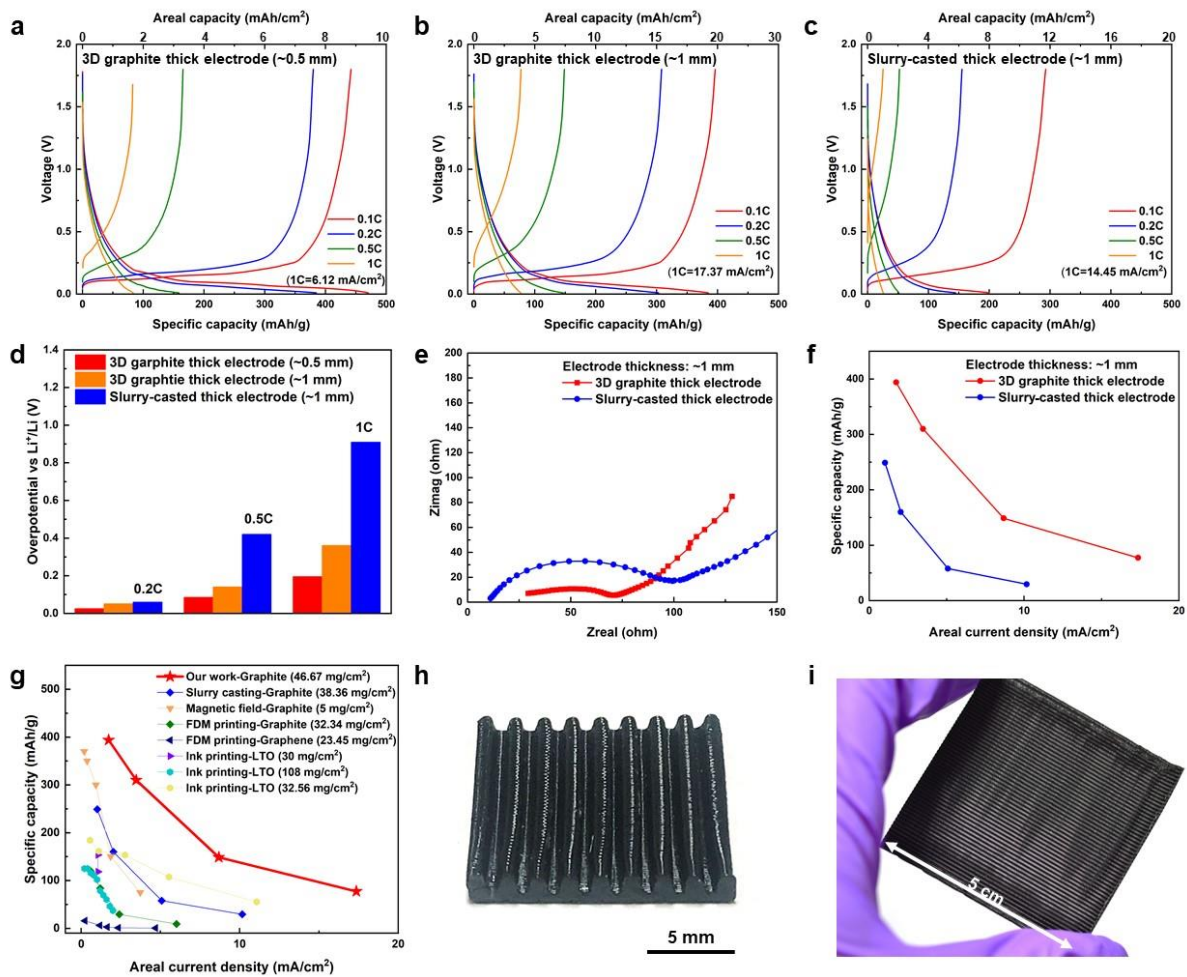

**Figure 5**. Electrochemical performances electrode and scalability and design flexibility of the 3D graphite manufactured with SEAM. (a and b) Galvanostatic cycling of 3D graphite thick electrodes with 0.5 mm and 1 mm thickness at rates of 0.2C, 0.5C, and 1C. (c) Galvanostatic cycling of the slurry-casted thick electrode with 1 mm thickness at rates of 0.2C, 0.5C, and 1C. (d) Overpotentials of the 3D graphite electrode and the slurry-casted thick electrode at different current densities. (e) EIS profiles of the 3D graphite electrodes and slurry-cased electrodes after 1C. (f) Comparison of the specific capacity of the 3D graphite electrode and the slurry-casted thick electrode at different areal current densities. (g) Comparison of the 3D graphite electrode and previously reported thick electrodes in the aspect to specific capacity as a function of areal current density. (h) Photo of 3D

graphite thick electrode with integrated structure. (i) Photo of large 3D graphite thick electrode showing scalability.

Figure 6 shows the potential of scale-up SEAM for structure electrodes. The electrode feedstocks were printed in one direction to have aligned structure and o the working size of the electrode and then attached on a current collector. The double-sided structure electrode is pressed and heated using a hot press to have a good between them. Scale-up SEAM allows for simple, safe, and high throughput fabrication of structure electrodes in practical use.

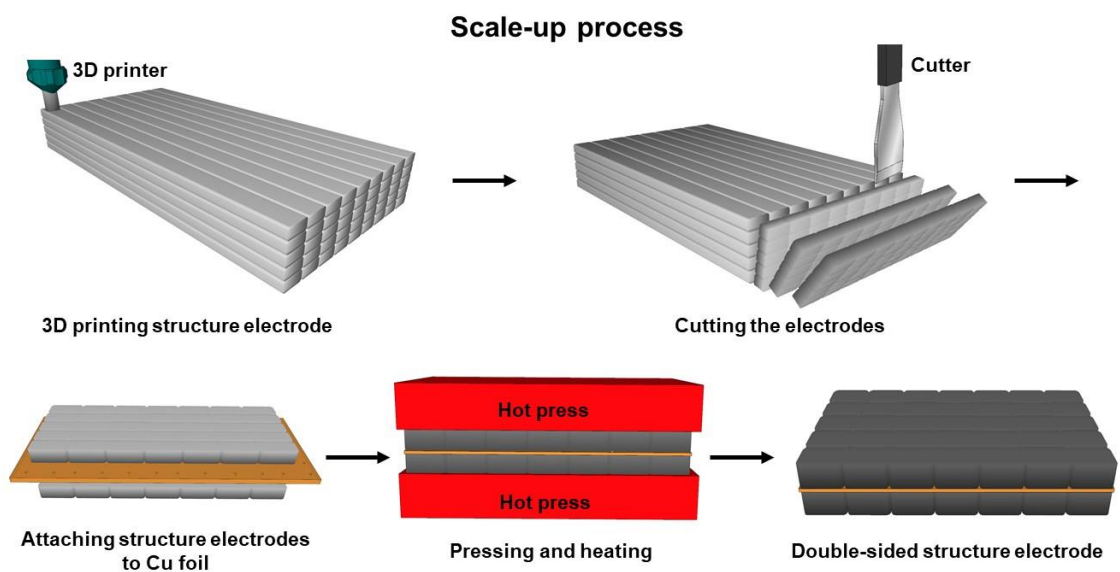

**Figure 6**. Scaling up design of SEAM for structured electrodes. The aligned electrodes are printed in one direction and cut into several pieces. After that, the electrodes are attached to the porous copper foil and pressed at high temperatures to make good contact between them. Double-sided structure electrodes can be fabricated fast and straightforward with SEAM.

## 3. Conclusion

We developed a new manufacturing route, called structure electrode additive manufacturing or SEAM, which controls the alignment of active materials in one direction to fabricate ultra-thick electrodes with aligned structures. The distinct advantage of our SEAM technique, including the use of high loading of graphite around 30wt.% as feedstocks, is that it can produce the structure electrodes with complex geometry in simple, safe, high throughput and scalable process. SHAPE technology using shear flow in SEAM allows for self-alignment of active materials, resulting in a low-tortuous arrangement. Aligned graphite provides a short path through the thickness, enabling rapid ion transport in impregnated liquid electrolyte. The 3D graphite electrode with high thickness ~ 1mm had a high active loading of 46.67 mg/cm$^2$ and exhibited a high through-thickness compressive strength of 7.1MPa, which is at least 30 times higher than that of 3D printed thick electrode made of graphite or graphene using ink-based printing. Our structure electrodes delivered a high specific capacity of 77.28 mAh/g and an areal capacity of 3.6 mAh/cm$^2$ at a current density of 17.36 mA/cm$^2$. This new manufacturing technique is prominent as a scalable manufacturing process, opening opportunities for practical energy storage applications.